\documentclass[journal]{IEEEtran}
\usepackage{amsmath,amssymb,amsfonts}  
\usepackage{booktabs}  
\usepackage{ragged2e}  
\usepackage{float}  
\usepackage{cite}  
\usepackage{multirow}  
\usepackage{hyperref}  
\usepackage{caption}  
\usepackage{subcaption}  
\usepackage{svg}  
\usepackage{algorithm}  
\usepackage{algpseudocode}  
\usepackage{enumitem}  
\usepackage{braket}  
\usepackage{array}  
\usepackage{caption}
\usepackage{graphicx}  
\graphicspath{{figs/}} 
\DeclareGraphicsExtensions{.jpg,.jpeg,.png,.PNG}  

\renewcommand{\arraystretch}{1.2}
\begin{document}

\title{SMSAT: A Multimodal Acoustic Dataset and Deep Contrastive Learning Framework for Affective and Physiological Modeling of Spiritual Meditation}

\author{Ahmad Suleman, Yazeed Alkhrijah, Misha Urooj Khan, Hareem Khan, Muhammad Abdullah Husnain Ali Faiz,  Mohamad A. Alawad, Zeeshan Kaleem,~\IEEEmembership{Senior Member,~IEEE}, Guan Gui,~\IEEEmembership{Fellow,~IEEE} 
\thanks{This work is supported and funded by the Deanship of Scientific Research at Imam Mohammad Ibn Saud Islamic University (IMSIU) grant number IMSIU-DDRSP2504. (*Corresponding author: Zeeshan Kaleem, e-mail: zeeshankaleem@gmail.com).

The source code and the dataset can accessed from here \url{https://www.kaggle.com/datasets/crdkhan/qmsat-dataset}.

Ahmad Suleman is with AITeC, National Center for Physics (NCP), Pakistan, and Vice-chairperson Community of Research and Development (CRD) (e-mail:ahmad.suleman@ncp.edu.pk)

Yazeed Alkhrijah and Mohamad A. Alawad are with the Department of Electrical Engineering, Imam Mohammad ibn Saud Islamic University (IMSIU), Saudi Arabia (e-mail: Ymalkhrijah@imamu.edu.sa, maawaad@imamu.edu.sa)

Misha Urooj Khan is with the European Organization for Nuclear Research, CERN, Switzerland, and chairperson of CRD (e-mail: misha.urooj.khan@cern.ch)

Hareem Khan and Muhammad Abdullah Husnain Ali Faiz are with the University of Engineering (UET), Taxila, and crew members CRD (e-mail: hareem.khan@students.uettaxila.edu.pk, 24-cp-50@students.uettaxila.edu.pk)

Zeeshan Kaleem is with the Department of Computer Engineering and the Interdisciplinary Research Center for Smart Mobility and Logistics, King Fahd University of Petroleum \& Minerals (KFUPM), Dhahran 31261, Saudi Arabia (e-mail: zeeshankaleem@gmail.com)

Guan Gui is with the College of Telecommunications and Information
Engineering, Nanjing University of Posts and Telecommunications, Nanjing
210003, China (e-mail: guiguan@njupt.edu.cn).}}

\markboth{IEEE Transactions on Affective Computing}%
{Shell \MakeLowercase{\textit{et al.}}: Your Title Here}

\maketitle

\begin{abstract}
Understanding how auditory stimuli influence emotional and physiological states is fundamental to advancing affective computing and mental health technologies. In this paper, we present a multimodal evaluation of the affective and physiological impacts of three auditory conditions, i.e., spiritual meditation (SM), music (M), and natural silence (NS), using a comprehensive suite of biometric signal measures. To facilitate this analysis, we introduce the Spiritual, Music, Silence Acoustic Time Series (SMSAT) dataset, a novel benchmark comprising acoustic time-series (ATS) signals recorded under controlled exposure protocols, with careful attention to demographic diversity and experimental consistency. 
To model the auditory-induced states, we develop a contrastive learning-based SMSAT audio encoder that extracts highly discriminative embeddings from ATS data, achieving 99.99\% classification accuracy in inter-class and intra-class evaluations. Furthermore, we propose the Calmness Analysis Model (CAM)—a deep learning framework integrating 25 handcrafted and learned features—for affective state classification across auditory conditions, attaining robust 99.99\% classification accuracy. In contrast, pairwise t-tests reveal significant deviations in cardiac response characteristics (CRC) between SM analysis via ANOVA inducing more significant physiological fluctuations.
Compared to existing state-of-the-art methods reporting accuracies up to 90\%, the proposed model demonstrates substantial performance gains (up to 99\%). This work contributes a validated multimodal dataset and a scalable deep learning framework for affective computing applications in stress monitoring, mental well-being, and therapeutic audio-based interventions.
\end{abstract}

\begin{IEEEkeywords}
Biomedical Signal Processing, Deep Learning, EEG, HRV, Spiritual Meditation, Stress Reduction.
\end{IEEEkeywords}

\section{Introduction}
Recent psychology, neuroscience, and biomedical engineering research has extensively explored how auditory stimuli influence cognitive and emotional states. Recently, it has been demonstrated that exposure to spiritual meditation (SM) audio stimuli, including recitations from holy texts, can effectively promote relaxation and enhance mental well-being \cite{palma2023stress, long2022wearable}. Researchers in \cite{shajari2023wearable, hemakom2024ecg, zawad2024stress} further investigated whether empirical techniques and machine learning methods, utilizing electroencephalography (EEG) signals, heart rate variability (HRV), and cortisol biomarkers, can objectively quantify the physiological and psychological effects of SM-based audio stimuli on stress reduction. This research is particularly significant given the World Health Organization's (WHO) report estimating that stress-related disorders constitute approximately 30\% of global health issues, impacting over 264 million individuals \cite{masri2023mental}.

Chronic stress significantly contributes to cardiovascular disease, depression, and cognitive decline, highlighting the urgent need for practical and noninvasive stress-reduction interventions. While general meditation and deep-breathing exercises are beneficial, the unique combination of rhythmic auditory stimuli and spiritual elements found in holy-text recitations positions SM audio as a promising tool for psychological rehabilitation. Researchers across psychology, neuroscience, and biomedical engineering have extensively studied how auditory stimuli influence cognitive and emotional states. Among various auditory interventions, SM audio, particularly religious book recitations, has effectively enhanced calmness and mental well-being. However, a critical research question remains regarding whether empirical methodologies can objectively quantify and validate these beneficial effects \cite{di2024anxiety, sabry2022healthcare}.
\vspace{-0.2 cm}
\subsection{Auditory Processing and Neural Mechanisms}
The human brain processes auditory stimuli through complex neural pathways, significantly influencing cognition and emotional states. Neuro-imaging research has demonstrated that specific auditory patterns can alter brainwave dynamics, hormone secretion, and neurotransmitter regulation. Functional magnetic resonance imaging (fMRI) and EEG research indicate that listening to structured rhythmic sounds, such as music or chanting, activates the limbic system—an essential neural region responsible for regulating emotions and memory formation \cite{almadhor2023efficient}. Brainwave activity is commonly classified into five distinct frequency bands. According to \cite{almadhor2023efficient}, theta (4–8 Hz) and alpha (8–12 Hz) wave activity increase significantly during SM, while the dominance of beta waves decreases. This suggests a shift toward a meditative and tranquil state, corroborating cognitive neuroscience research that theta waves are linked to increased attentional concentration and reduced tension.
\vspace{-0.4 cm}
\subsection{Physiological Effects of Spiritual Meditation-based Audio Stimuli}
Audio stimuli such as spiritual meditation containing chanting, gongs, binaural beats, natural sounds, Tibetan singing bowls, and recitations of holy texts affect neurological and circulatory systems. When people interact with these auditory stimuli, they experience a decrease in stress hormones such as cortisol and a transition in brain wave patterns towards more relaxed states, including alpha and theta waves. These physiological responses diminish heart rate and blood pressure, thus leading to a mood of balance and tranquility. Regular participation in these auditory activities leads to enhanced emotional regulation, strengthened immune function, and a more focused and balanced life, suggesting a substantial link between sound, meditation, and physical well-being.

Several real-world studies have examined how SM reduces physiological stress \cite{agarwal2022}.  A review of peer-reviewed studies conducted between 2010 and 2022 found that listening to SM lowered cortisol levels by an average of 35\% compared to control groups exposed to quiet or non-religious audio stimuli \cite{Azzeh2024}. Cortisol is the primary stress hormone. In a controlled trial by \cite{Lubab2024}, HRV and autonomic nerve system stability were measured for 120 volunteers who listened to the SM audio. Analysis of skin conductance response (SCR), a physiological indicator of stress, revealed that participants exposed to SM audio for twenty minutes experienced a 42\% reduction in SCR levels, signifying a calming effect on the nervous system \cite{Muhith2024}. Additionally, previous research \cite{nayef2018effect} demonstrated that SM enhances theta wave activity, a brainwave pattern closely linked to deep relaxation and reduced anxiety.

The authors in\cite{sadeghi2023study} demonstrated that SM significantly enhanced HRV, reflecting improved parasympathetic nervous system activity. Additionally, galvanic skin response (GSR)—a measure of skin conductance that mirrors autonomic nervous system arousal \cite{hanafi2024effect}—shows reduced values following exposure to SM audio stimuli, indicating lowered stress levels.

\begin{table*}[h]
\centering
\caption{Overview of Datasets with Data Collection Types and Availability.}
\label{tab:datasets_overview}
\resizebox{\textwidth}{!}{%
\begin{tabular}{l l l l}
    \toprule
    \textbf{Dataset} & \textbf{Description} & \textbf{Data Collected} & \textbf{Availability} \\
    \midrule
    \textbf{DEAP} \cite{Koelstra2012} & Emotion recognition using physiological signals such as EEG and ECG for analyzing emotional responses to stimuli. & EEG, ECG, GSR & Public Access \\
    \textbf{ForDigitStress} \cite{aditya2023} & Multi-modal dataset aimed at stress detection from biosignals like EEG and ECG, focusing on automatic stress recognition. & EEG, ECG, EDA & Limited Access \\
    \textbf{AMIGOS} \cite{Correa2018} & Dataset designed for mood, personality, and affect research using multimodal measurements such as EEG and EDA. & EEG, EDA & Public Access \\
    \textbf{DREAMER} \cite{Katsigiannis2017} & Emotion recognition through EEG and ECG signals, focusing on multimodal signal integration for emotion analysis. & EEG, ECG & Public Access \\
    \textbf{TAFFS} \cite{masri2023mental} & A dataset designed for detecting mental stress in the workplace using physiological and behavioral signals. & Heart Rate, EDA, ECG & Public Access \\
    \textbf{Biosignal Stress Dataset} \cite{Yadav2023} & Focuses on stress detection using physiological signals such as EEG, ECG, and GSR. & EEG, ECG, GSR & Public Access \\
    \textbf{PHQ-9} \cite{Muhith2024} & A dataset for depression recognition using self-reported data and physiological signals such as EEG. & EEG, self-reported & Limited Access \\
    \textbf{EmotionEEG} \cite{palma2023stress} & A dataset for emotion recognition using EEG signals in response to various emotional stimuli. & EEG & Public Access \\
    \textbf{StressLog} \cite{zawad2024stress} & This dataset involves stress detection from wearable sensors, focusing on HRV and other physiological features. & HRV & Public Access \\
    \textbf{WearableStress} \cite{shajari2023wearable} & Dataset aimed at analyzing mental stress and emotion using data collected from wearable devices, such as EEG and ECG. & EEG, ECG, EDA & Public Access \\
    \textbf{Emotion Recognition Dataset} \cite{long2022wearable} & Multi-modal dataset for emotion recognition using physiological signals like EEG and ECG, targeting real-time emotion analysis. & EEG, ECG & Limited Access \\
    \textbf{UANet} \cite{di2024anxiety} & Dataset used to detect anxiety and other mental health conditions from physiological signals, including ECG and EEG. & ECG, EEG & Limited Access \\
    \textbf{WearableAnxiety} \cite{sabry2022healthcare} & Focused on real-time anxiety detection from physiological signals like ECG and EEG. & ECG, EEG & Limited Access \\
    \textbf{MentalHealthECG} \cite{gedam2021} & A dataset that recognizes mental health conditions, particularly stress and anxiety, from ECG signals. & ECG & Public Access \\
    \textbf{StressWearables} \cite{bhuthegowda2024} & This dataset includes wearable data for automatic stress recognition, focusing on HRV and other physiological markers. & Heart Rate, GSR, ECG & Public Access \\
    \textbf{YogaStress} \cite{agarwal2022} & A dataset for studying the effects of yoga practices on mental health through sensors like EEG. & EEG, EDA & Limited Access \\
    \textbf{QuranEmotion} \cite{sadeghi2023study} & Dataset focused on analyzing the emotional responses during Quran recitation through physiological signals like EEG and heart rate variability. & EEG, HRV & Public Access \\
    \bottomrule
\end{tabular}%
}
\end{table*}

\begin{figure*}[hbt!]
    \centering
    \includegraphics[width=0.99\textwidth]{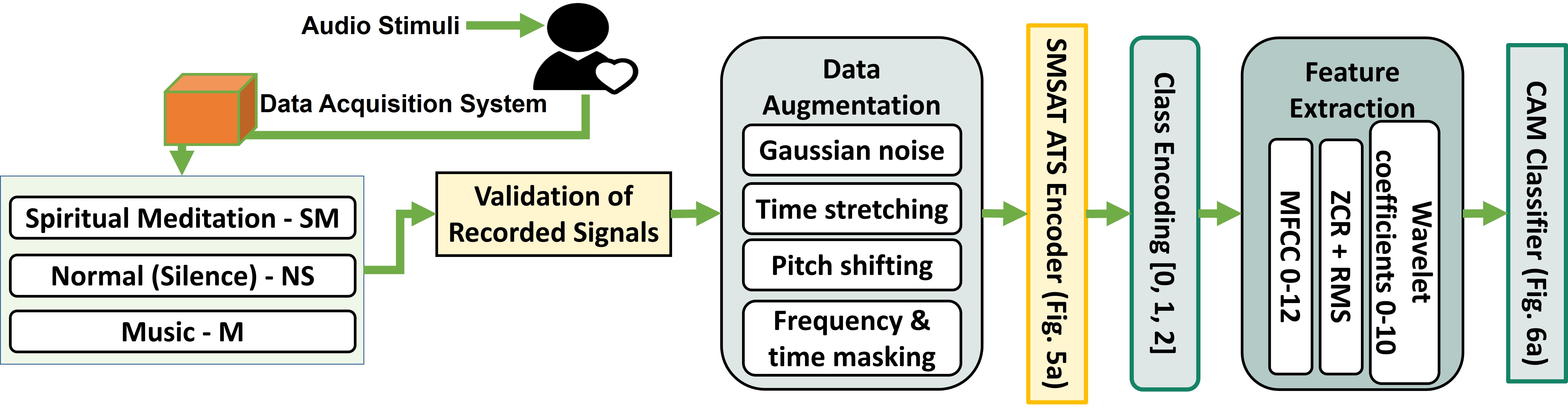} 
    \caption{Detailed flow graph of the proposed methodology.}
    \label{fig:meth}
    \vspace{-0.6 cm}
\end{figure*}  

\vspace{-0.4 cm}
\subsection{Datasets, Limitations \& Experimental Analysis}
Various datasets have been assembled to examine the physiological and neurological impacts of auditory stimuli \cite{lokare2024}. While these datasets provide a valuable starting point, there are notable limitations. The SM EEG dataset is not publicly accessible, limiting research reproducibility. In Table \ref{tab:datasets_overview}, an overview of the datasets used for stress detection and related applications is provided, and the limitations of existing datasets are highlighted. Some of those are discussed here as well. There is a lack of Quran-specific public datasets; however, DEAP \cite{Koelstra2012} and AMIGOS \cite{Correa2018} collect EEG, but they do not specifically gauge reactions to QR as in \cite{nayef2018effect}. Also, there is problem of small sample size. Many datasets for stress detection \cite{palma2023stress}, contain a limited number of participants (\textless100), reducing the generalization of results. Moreover, the experimental conditions were not the same, which resulted in differences in electrode placements and device configurations, which in turn introduced discrepancies in results \cite{hemakom2024ecg, Al-Qazzaz2015, Katsigiannis2017}. 

The other major limiting factor is the limited access to datasets. For example, the datasets like \cite{long2022wearable, hemakom2024ecg, masri2023mental, almadhor2023efficient, Lubab2024, Yadav2023, nayef2018effect,  hanafi2024effect, cosic2024, aditya2023, ayyala2024, lokare2024} are restricted in terms of access, thus limiting the ability of researchers to reproduce and compare with their results. The other lack lies in the provision of single modalities without incorporating other signals like HRV \cite{zawad2024stress}, GSR \cite{masri2023mental}, or EEG \cite{Koelstra2012} for a more holistic analysis \cite{hemakom2024ecg}, \cite{masri2023mental}, \cite{zheng2015}, \cite{nayef2018effect},  \cite{hanafi2024effect}, \cite{cosic2024}, \cite{ayyala2024}. Last but not least, there is a lack of diverse population representation in the existing datasets. For instance, the AI-based stress detection datasets are based on a homogeneous group (e.g., university students), which limits their applicability to diverse populations.  
\vspace{-0.4 cm}
\subsection{AI-Based Calmness/ Stress Detection Models}
Various sensors capture signals for stress and calmness evaluation \cite{rojas2024bioinformatics, patel2021ann, patel2021ann} like electronic stethoscopes,  MEMS microphones, piezoelectric sensors, and multi-modal sensors.  Various wearable devices equipped with sensors to measure heart rate, electrodermal activity, EEG, and electrocardiogram (ECG) have been integrated with machine learning (ML) and deep learning (DL) algorithms to create models that accurately detect stress levels. In \cite{palma2023stress}, the authors used ML and DL for stress detection using the Empatica E4 bracelet, which captures physiological data such as heart rate and GSR, with 85\% accuracy in stress response detection.  In \cite{Al-Qazzaz2015}, the authors emphasized that AI-based music therapy have the potential to offer personalized mental management \cite{hemakom2024ecg}. Gender and hormonal phases can influence stress responses, with ML models showing a 92\% accuracy in distinguishing between stress levels in women across different menstrual phases and for me,n providing a basis for developing more individualized stress detection systems. 

In \cite{di2024anxiety}, AI detected anxiety by analyzing variations in heart rate and skin conductance with 90\% accuracy.  In \cite{gedam2021}, the authors developed models to assess mental stress in clinical and non-clinical settings, with 85\% to 95\% accuracy. In conclusion, integrating AI with sensors can revolutionize how stress and calmness are detected, allowing for real-time monitoring and personalized interventions. 
\vspace{-0.4cm}
\subsection{Paper Contributions}
To address the identified research gaps in evaluating the physiological and emotional impact of auditory stimuli, this paper makes the following key contributions, as illustrated in Fig. 1:

\subsubsection{Creation of a Novel Multimodal Dataset (SMSAT)}
We introduce the Spiritual, Music, Silence Acoustic Time Series (SMSAT) dataset, which comprises acoustic time-series (ATS) signals recorded under three auditory conditions: spiritual meditation (SM), music (M), and natural silence (NS). The dataset is designed with demographic diversity and experimental rigor in mind, and it is made publicly available to promote reproducibility and further research.

\subsubsection{Rigorous Dataset Validation via Signal Processing Techniques}
We employ Hilbert Transform-based envelope extraction to verify signal integrity and use Root Mean Square Error (RMSE) metrics to quantify deviations between recorded and theoretical signal envelopes. Additionally, FFT-based spectral analysis is conducted to validate the fidelity of the recording equipment in capturing critical frequency and acoustic characteristics.

\subsubsection{First Quantitative Framework for ATS Signal Validation}
To the best of our knowledge, this is the first work to present a quantitative validation framework for acoustic time-series datasets based on theoretical signal modeling, establishing a reproducible benchmark for dataset credibility and sensor accuracy in affective computing research.

\subsubsection{Design of the SMSAT ATS Encoder with Contrastive Learning}
We propose a customized contrastive learning-based encoder to generate discriminative and class-specific embeddings from ATS data. The encoder supports a large-scale model architecture of 11.23 million parameters and approximately 200 million FLOPs, ensuring scalability and computational efficiency.

\subsubsection{Development of the Calmness Analysis Model (CAM)}
We design a deep learning classification model integrating 25 handcrafted and learned features to distinguish affective states induced by different auditory conditions. The CAM achieves near-perfect classification accuracy across multiple test scenarios.

\subsubsection{Comprehensive Statistical and Visualization Analysis}
Using ANOVA and pairwise t-tests, we demonstrate that SM elicits physiological responses closely aligned with NS, while significantly differing from M. Visualization tools such as t-SNE plots and heatmaps further support the separability and consistency of the induced affective states.

\section{Proposed SMSAT (Spiritual, Music, Silence Acoustic Time Series) Dataset}
ATS signals provide critical information about the person's physiological state and human organ (cardiac activity) under consideration. We propose a new dataset of ATS-recorded signals under different auditory stimuli conditions to analyze the impact on CRC.  Mathematically, the dataset can be represented as:
\vspace{-0.2 cm}
\begin{equation}
D = \{(ats_{n}(t), y_n), (ats_{m}(t), y_m), (ats_{s_m}(t), y_{s_m})\}_{t=1}^{T},
\end{equation}
where $ats_n (t)$, $ats_m (t)$, and $ats_{s_m} (t)$ denote time-series ATS signals, and $y_n$, $y_m$, and $y_{s_m}$ represent their corresponding class labels: $y \in \{ \text{Normal (Silence)}, \text{Music}, \text{Spiritual Meditation} \}$.
\vspace{-0.4 cm}
\subsection{Data Collection and Acquisition Device}
The ATS signals were recorded in a controlled environment using a custom-designed acquisition system. The experimental setup consists of: a stethoscope $H$ for auscultation, high-sensitivity microphone $M$ for capturing vibrations, a sound card $SC$ for analog-to-digital conversion, and a laptop $L$ for data storage in \texttt{.wav} format. Fig.~\ref{fig:setup1}  (a) shows the acquisition device setups with key devices named there, where each subject remained seated in a rest position during recording. Signals are collected on alternative days, ensuring a natural variation in physiological responses. 
\vspace{-0.4 cm}
\subsection{Demographic Information}
\begin{figure}[h]
\vspace{-0.3 cm}
    \centering
    (a) \includegraphics[width=.45\textwidth]{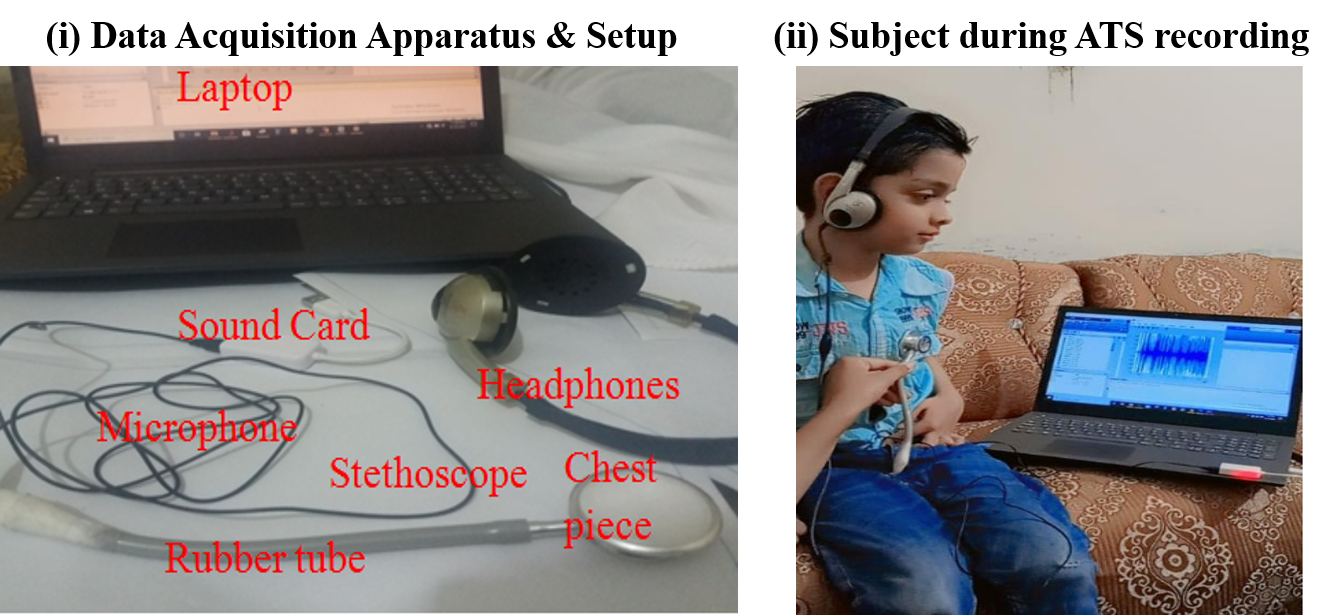} \\
    (b) \includegraphics[width=.45\textwidth]{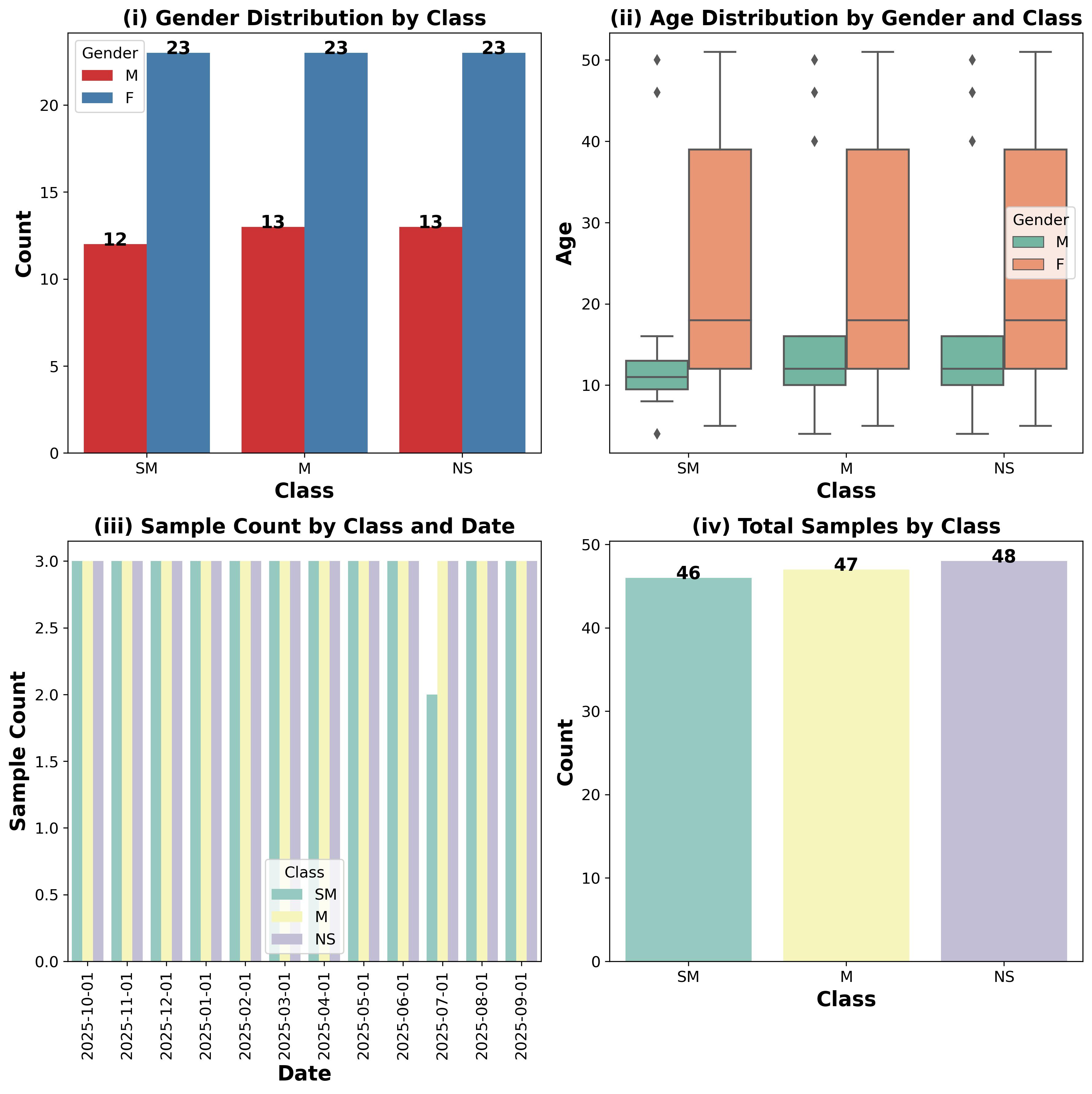} \\
    (c) \includegraphics[width=.46\textwidth]{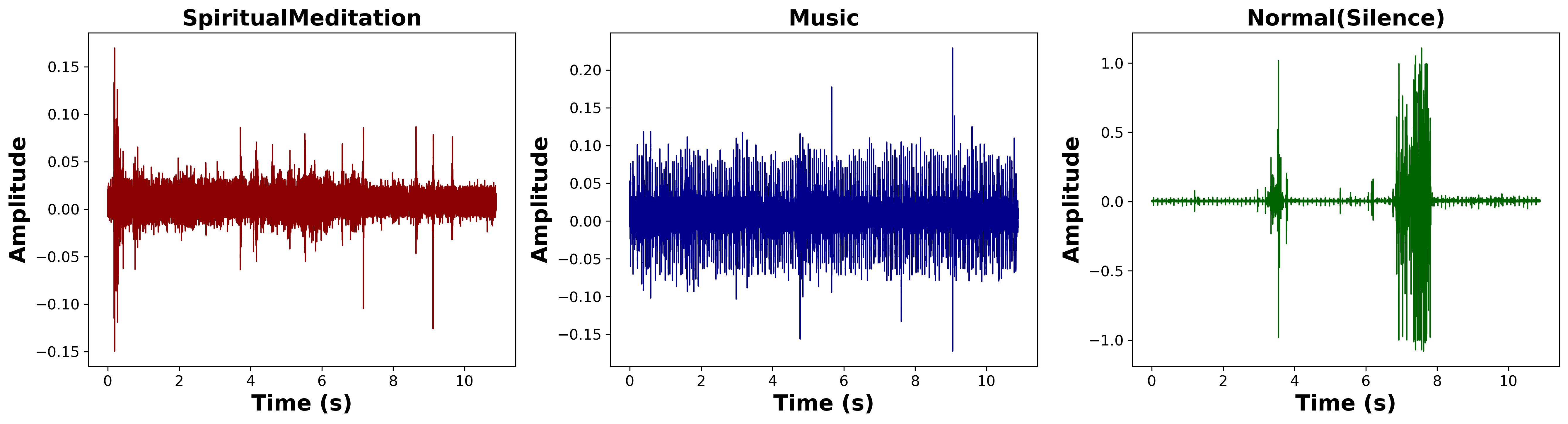}    
    \caption{(a) Data setup (b) SMSAT Statistics (c) Time domain plot.}
    \label{fig:setup1}
\vspace{-0.5 cm}
\end{figure}

The dataset contains 141 ATS recordings evenly distributed across three categories.  Each ATS signal is recorded for 60 seconds and stored in \texttt{.wav} format with a sampling rate of 16 kHz. The dataset is diverse in age (3 to 55 years) and gender-balanced, meaning both male and female subjects are under experimentation in the study, thus ensuring fair representation across different demographic groups as shown in Fig. \ref{fig:setup1} (b).
\vspace{-0.4 cm}
\subsection{Experimental Conditions}
\subsubsection{Normal/Silence Sitting State –\( ats_n(t) \)} In this condition, the subject is seated in a quiet environment. The baseline, or resting state, is established in which ATS is recorded without external auditory stimulation. This state serves as a control measurement, providing a baseline against which changes during other auditory conditions can be compared. Furthermore, any detected variations in ATS under experimental settings should be attributed to particular aural stimuli rather than surrounding or outside variables. This quiet state mitigates external effects, such as ambient noise or auditory distractions, ensuring that external noises do not affect the subject's physiological reactions. Furthermore, this condition enabled us to calibrate the dataset, determining the subject's baseline profile of ATS before the introduction of any auditory stimuli.

\subsubsection{Listening to Music – \( ats_m(t) \)} The participant in this experimental condition listens to instrumental music while the ATS captures data. We select instrumental music, devoid of lyrics, to reduce the cognitive processes linked to language. This auditory stimulus provoked emotional reactions and cognitive alterations that influenced the subject's physiological metrics. The reactions included variations in heart rate, changes in relaxation levels, and shifts in concentration. The music's rhythm, harmony, and melody also influenced the temporal variations in ATS, resulting in synchronization effects and distinct patterns.

\subsubsection{Listening to Spiritual Meditation Audio – \( ats_{s_m}(t) \)} During ATS signal recording, the subject listened to spiritual meditation-based audio content. These audios included guided meditation cues, ambient sounds, or spiritual guides like listening to holy texts to induce a meditative state. This condition is specifically included to examine the effects of meditative or spiritual audio content on the subject's physiological and temporal responses, as captured by ATS. Subjects said this reduced stress levels, induced relaxation, and increased parasympathetic activity.

\vspace{-0.4cm}
\subsection{Recorded Dataset Validation}
The recorded dataset is analyzed mathematically to confirm the signals are consistent with their theoretical expectations. The primary method involves using the hilbert transform (HT) to extract the envelope of the recorded signals \cite{rojas2024bioinformatics}. The envelope of the signal is computed as

\begin{equation}
A(t) = \sqrt{x^2 (t) + H^2 [x(t)]},
\end{equation}
where \( x(t) \) is the raw recorded ATS signal and \( H[x(t)] \) is the HT. From this envelope, the theoretical approximation of the recorded signal is computed as
\begin{equation}
x_{\text{theo}} (t) = A(t) \cos(2\pi f_c t + \phi),
\end{equation}
where \( f_c \) is the characteristic frequency of the class, selected based on empirical observations \cite{sadeghi2023study}

\begin{equation}
f_c =
\begin{cases}
25 \text{ Hz,} & \text{Spiritual Meditation} \\
20 \text{ Hz,} & \text{Music}. \\
30 \text{ Hz,} & \text{Normal (Silence)}
\end{cases}
\end{equation}

\begin{figure*}
\vspace{-0.5cm}
    \centering
    \includegraphics[width=0.8\textwidth]{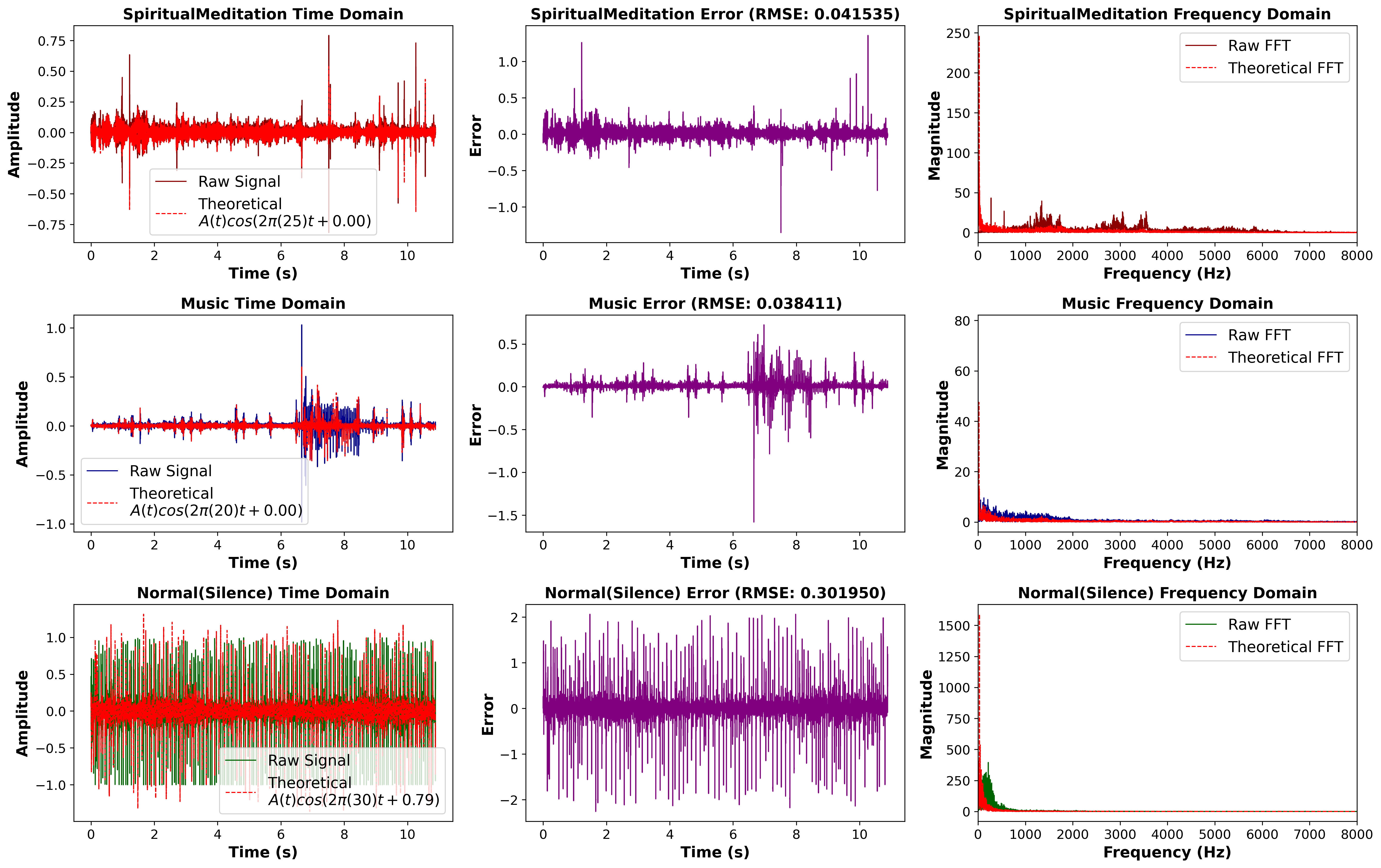} 
    \caption{Time domain and FFT comparison of raw and theoretical ATS signals.}
    \label{fig:pcg_signals1}
\end{figure*}
To measure the similarity between the recorded signal and its theoretical counterpart, RMSE is computed \cite{Yadav2023} as

\begin{equation}
\text{RMSE} = \sqrt{\frac{1}{N} \sum_{t=1}^{N} (x(t) - x_{\text{theo}} (t))^2}.
\end{equation}
A lower RMSE value indicates that the theoretical model accurately represents the recorded signal. The computed RMSE values for each class are $: \text{RMSE}_{\text{SM}} = 0.0269, \quad \text{RMSE}_{\text{Music}} = 0.0902, \quad \text{RMSE}_{\text{Normal}} = 0.0084.$ These results indicate that the SM and NS ATS signals match strongly with their theoretical models, while M signals show more variability. FFT is applied to analyze the spectral content of the signals, and its magnitude spectrum is computed as
\begin{equation}
|X(f)| = \sqrt{\text{Re}(X(f))^2 + \text{Im}(X(f))^2}.
\end{equation}

A comparison of the recorded FFT spectrum and the theoretical spectrum reveals that SM ATS signals exhibit distinct low-frequency patterns around \( 25 \) Hz. M ATS has a broader frequency range with more fluctuations. NS ATS shows minimal energy content across frequencies, validating the device's ability to detect silence. This analysis validates the recorded dataset in several key aspects:
\begin{enumerate}
    \item The envelope extraction and theoretical modeling confirm that the recordings follow expected amplitude modulations.
    \item The RMSE values indicate strong alignment between the recorded and expected signals, particularly for spiritual meditation and silent data.
    \item The FFT comparison ensures that spectral components are correctly captured, verifying the reliability of the recording setup.
\end{enumerate}
The ATS signal analysis reveals apparent differences in CRC pattern across auditory stimuli in Fig. ~\ref{fig:pcg_signals1}.  RMSE, which measures the difference between the actual and theoretical signals, is lowest for NS(0.0084), indicating stable CRC. Moderate variation suggests a structured but slightly varied pattern for SM(0.0262). M(0.0884) has the highest RMSE, indicating high CRC irregularities. M(0.0537) has the highest mean envelope amplitude, followed by SM(0.0177) and NS(0.0063). A higher mean amplitude suggests more significant variations in ATS, with M having the most pronounced effects. The standard deviation of the envelope is highest for M(0.0686) and lowest for NS(0.0047), showing that M leads to the most fluctuations in CRC while SM has a more balanced response. Energy levels also reinforced these trends, as M(664.36) has the highest CRC, SM(60.95) is moderate, and NS(7.13) is resting. These findings imply that SM calms while M increases CRC variability and excitement. These results and Fig. \ref{fig:pcg_signals1} confirm that the designed device correctly captures SM, music, and silent environments, ensuring the collected dataset is scientifically valid for further analysis.
\vspace{-0.2cm}
\section{Proposed SMSAT ATS Encoder}
Before passing raw data, some steps are adopted, like augmenting ATS signals 5 times. Then, their relevant spectrograms are generated and passed to a custom encoder for embedding learning. 
\vspace{-0.5 cm}
\subsection{Data Augmentation Techniques}
\begin{figure}[hbt!]
    \centering
    \includegraphics[width=0.5\textwidth]{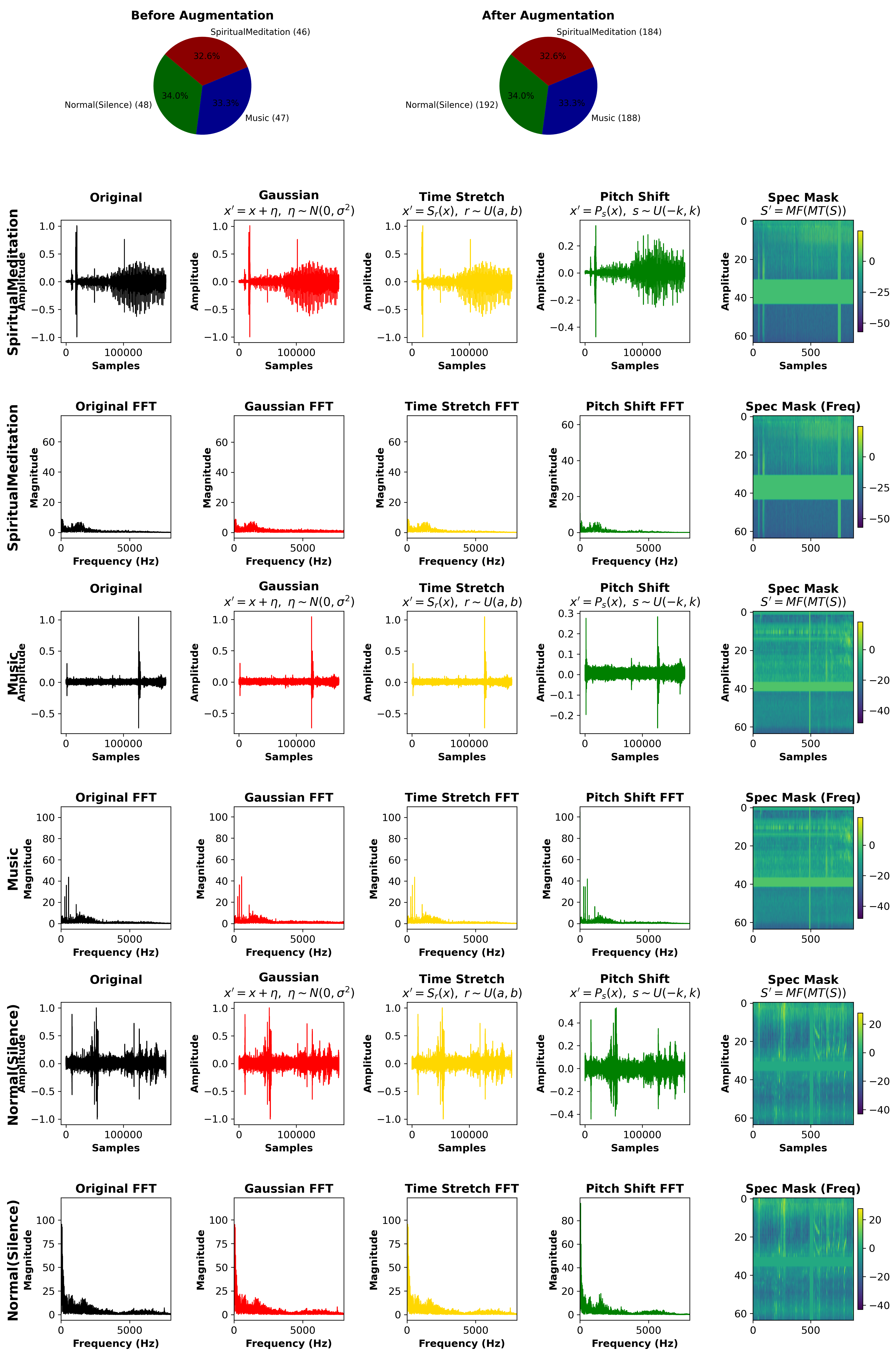}
    \caption{Data augmentation techniques applied to ATS signals.}
    \label{fig:data_augmentation}
\vspace{-0.3 cm}
\end{figure}

Fig. \ref{fig:data_augmentation} illustrates the data augmentation techniques applied to ATS signals like gaussian noise, time stretching, pitch shifting, and spectrogram masking. Those techniques are elaborated here in detail.

\subsubsection{Additive gaussian noise} adds random noise to simulate real-world environments. Mathematically, it can be expressed as
\begin{equation}
    x' = x + \eta, \quad \eta \sim \mathcal{N}(0, \sigma^2),
\end{equation}
where \( x' \) is the noisy signal, \( x \) is the original signal, \( \eta \) is the additive noise and \( \mathcal{N}(0, \sigma^2) \) is a gaussian distribution with mean 0 and variance \( \sigma^2 \). This technique improves noise robustness in ATS signals.

\subsubsection{Time stretching} randomly alters the speed of the audio without changing the pitch to simulate variations in speed. Mathematically it can be expressed as
\begin{equation}
    x' = S_r(x), \quad r \sim U(a,b),
\end{equation}
where \( x' \) is the time-stretched signal, \( x \) is the original signal, \( S_r(x) \) represents the time-stretched signal based on a stretching factor \( r \) and \( r \sim U(a,b) \) is uniformly distributed between \( a \) and \( b \).

\subsubsection{Pitch shifting} shifts the pitch randomly within a given range to ensure pitch invariance. Mathematically it can be expressed as
\begin{equation}
    x' = P_s(x), \quad s \sim U(-k,k),
\end{equation}
where \( x' \) is the pitch-shifted signal, \( x \) is the original signal, \( P_s(x) \) represents the pitch-shifted signal with a shift factor \( s \), \( s \sim U(-k,k) \) is uniformly distributed between \( -k \) and \( k \), representing the pitch shift range.

\subsubsection{Spectrogram frequency \& Time masking} makes the model robust to missing or occluded frequencies. It can be expressed as
\begin{equation}
    S' = M_F(M_T(S)),
\end{equation}
where \( S' \) is the masked spectrogram, \( S \) is the original spectrogram, \( M_T(S) \) applies time masking to the spectrogram, and \( M_F(S) \) applies frequency masking to the spectrogram. 

\vspace{-0.3 cm}
\subsection{Melspectrogram Generation}
Embedding learning tasks necessitate the extraction of robust feature representations. Self-supervised contrastive learning is applied to learn meaningful embeddings directly from the data, eliminating the need for manual labeling \cite{almadhor2023efficient, Yadav2023}. Let $x \in \mathbb{R}^T$ be a discrete-time ATS signal of length $T$, sampled at a rate $f_s$. Our goal is to learn a mapping function that transforms this ATS into a low-dimensional latent space representation as
\begin{equation}
    f_\theta: \mathbb{R}^T \rightarrow \mathbb{R}^d, \quad d \ll T.
\end{equation}

A raw ATS $x(t)$ contains temporal variations but lacks explicit frequency information. To capture both time and frequency features, we use the short-time Fourier transform (STFT), which transforms the signal into the frequency domain as
\begin{equation}
    X(f,t) = \sum_{n=0}^{N-1} x(n) w(n-t) e^{j 2\pi f n},
\end{equation}
where $X(f,t)$ is the complex-valued frequency representation, $w(n)$ is a window function, $N$ is the window length, $f$ represents frequency, and $t$ represents time. The spectrogram of the signals is then computed as
\begin{equation}
    S(f,t) = |X(f,t)|^2.
\end{equation}

The Mel filter scale is applied to generate frequency banks that align with a perceptually relevant scale. 
\begin{equation}
    S_{\text{mel}}(k,t) = \sum_{f} H_k (f) S(f,t),
\end{equation}
where $H_k(f)$ are the Mel filter weights. The Mel-scale frequency mapping is defined as
\begin{equation}
    m(f) = 2595 \log_{10} (1 + f/700).
\end{equation}
\vspace{-0.6 cm}
\subsection{Proposed Model Architecture}
The Mel-spectrogram $S_{\text{mel}} \in \mathbb{R}^{F \times T}$ is now passed through an SMSAT encoder to learn hierarchical representations. The network consists of convolutional layers with residual connections to extract robust features. The SMSAT encoder is mathematically formulated as with the first convolutional layer is represented as
\begin{equation}
    h_0 = \sigma(W_0 * S_{\text{mel}} + b_0),
\end{equation}
where $W_0 \in \mathbb{R}^{64 \times 1 \times 7 \times 7}$. For each residual block
\begin{equation}
    h_i = \sigma(W_i * h_{i-1} + b_i),
\end{equation}
where \( h_i \) represents the output of the \( i \)-th residual block, \( W_i \) is the weight matrix at the \( i \)-th layer, \( h_{i-1} \) is the input from the previous residual block, \( b_i \) is the bias term for the \( i \)-th layer, \( \sigma \) denotes an activation function (e.g., ReLU),
and \( i \) represents different residual layers. Residual connections ensure effective gradient flow, addressing the vanishing gradient problem.  Global average pooling (GAP) is applied to aggregate high-level features while preserving essential information. This results in a compressed representation as
\begin{equation}
    z = \frac{1}{HW} \sum_{i=1}^{H} \sum_{j=1}^{W} h_l(i,j),
\end{equation}
where \( H \) and \( W \) are the height and width of the feature map, respectively. \( h_l(i,j) \) represents the feature map at location \( (i,j) \) and \( z \) is the compressed feature representation after applying GAP. A fully connected projection head maps features into a latent space as
\begin{equation}
    p = W_p z + b_p, \quad W_p \in \mathbb{R}^{d \times 512},
\end{equation}
where \( p \) is the projected latent vector, \( W_p \) is the projection matrix with dimensions \( d \times 512 \) and \( b_p \) is the bias term for the projection layer.

\begin{figure*}[h]
\vspace{-0.5 cm}
    \centering
    (a)\includegraphics[width=0.98\textwidth]{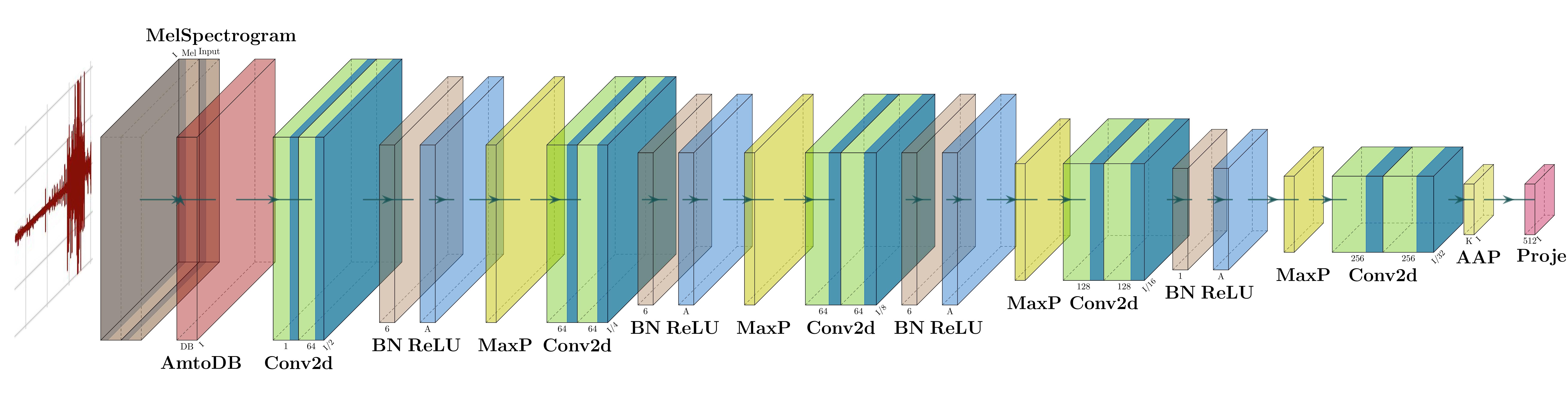} \\
    (b)\includegraphics[width=.75\textwidth]{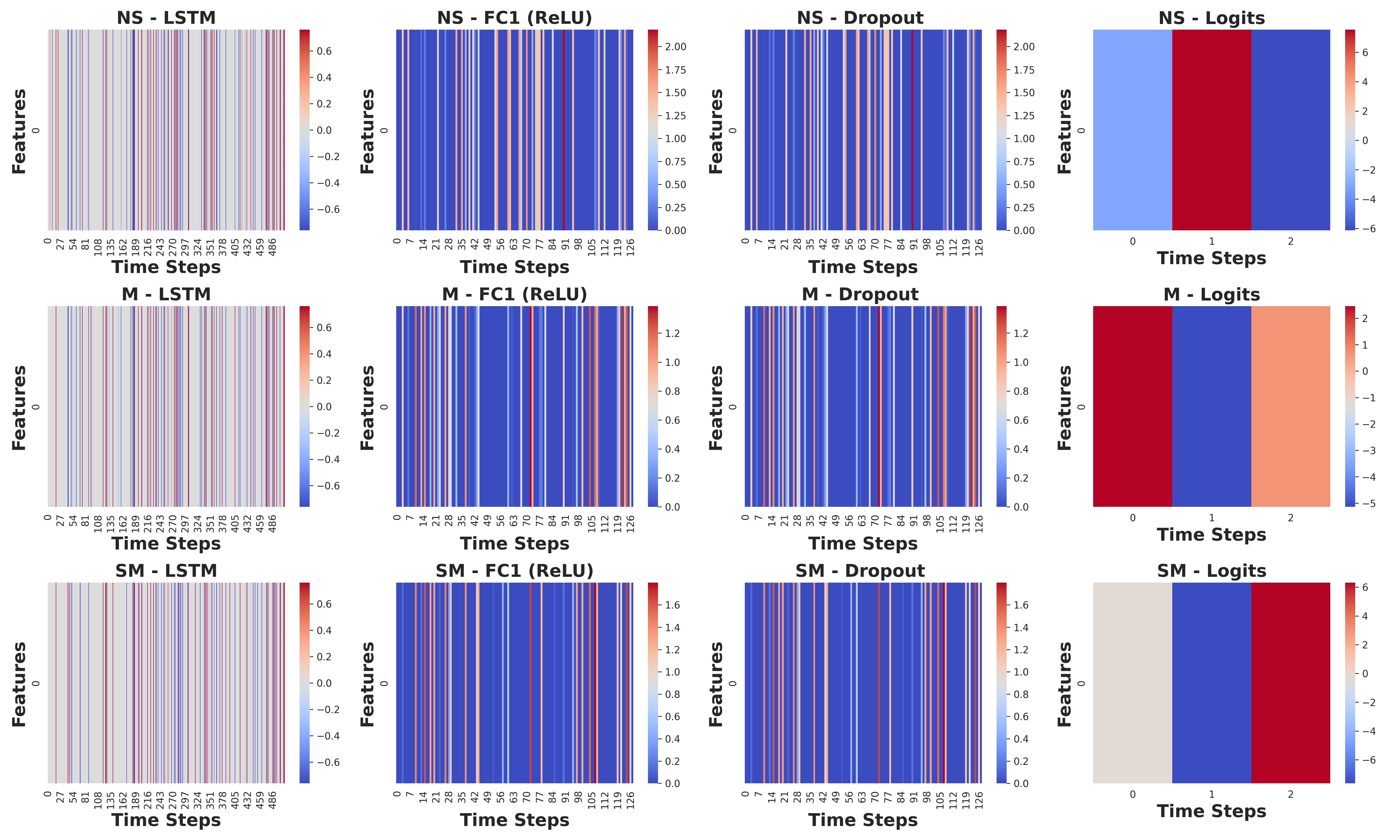} 
    \caption{SMSAT ATS encoder (a) Architecture (b) Visualization of layers for ATS signals.}
    \label{fig:smsatmodel}
\vspace{-0.6 cm}
\end{figure*}

SMSAT encoder is trained using a special contrastive loss function, which minimizes the distance between embeddings of augmented versions of the same sample.
\begin{equation}
    \mathcal{L} = \frac{1}{N} \sum_{i=1}^{N} || f_\theta(x_1^{'}) - f_\theta(x_2^{'}) ||^2,
\end{equation}
where \( \mathcal{L} \) is the contrastive loss, \( N \) is the number of samples in the batch, and \( f_\theta(x_1^{'}) \) and \( f_\theta(x_2^{'}) \) are the embeddings of the augmented versions of the same sample \( x_1^{'} \) and \( x_2^{'} \), respectively. \( \theta \) represents the parameters of the SMSAT encoder.The detailed architecture of the proposed SMSAT ATS encoder is shown in Fig.~\ref{fig:smsatmodel}(a), while the transformations at each layer are depicted in Fig.~\ref{fig:smsatmodel}(b).
\vspace{-0.3 cm}

\begin{figure*}[h]
\vspace{-0.5 cm}
    \centering
    (a) \includegraphics[width=0.65\textwidth]{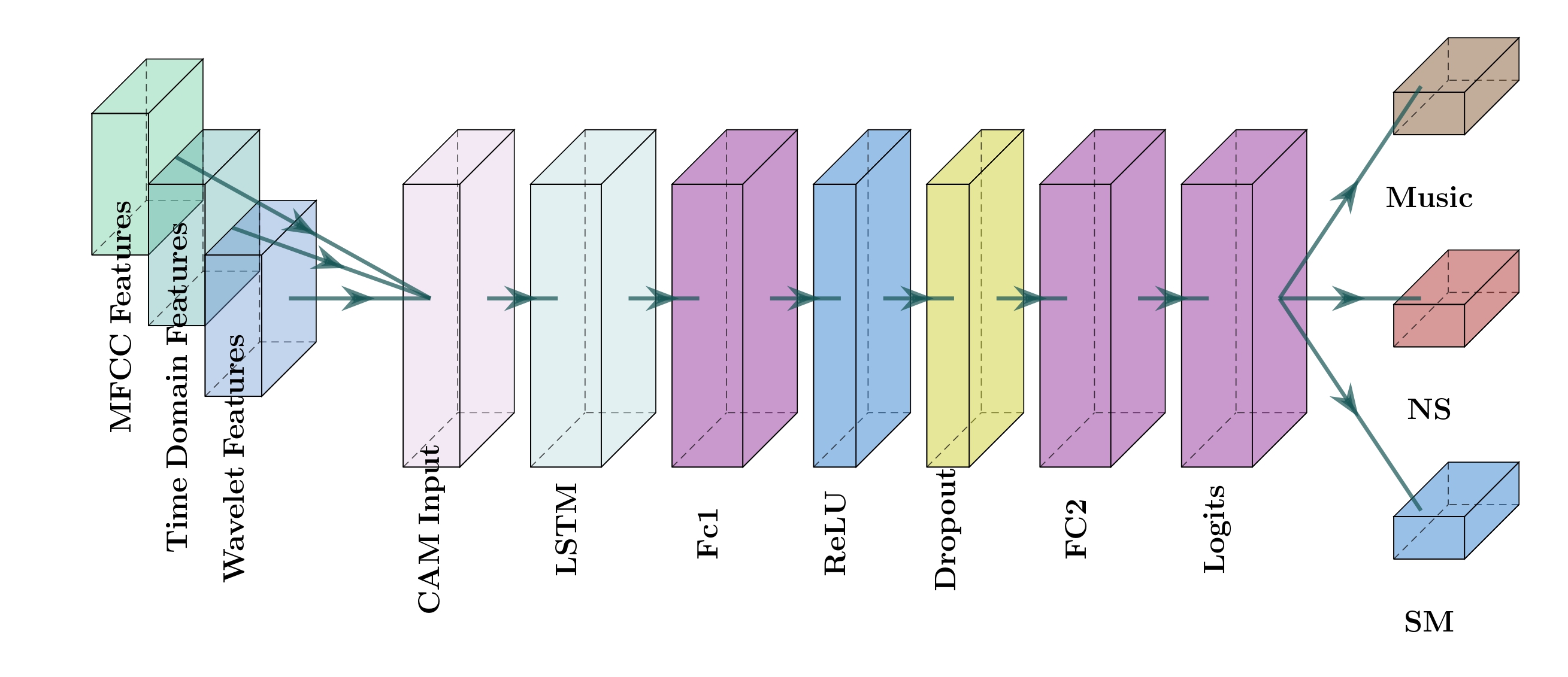} \\
    (b-i)\includegraphics[width=0.96\textwidth]{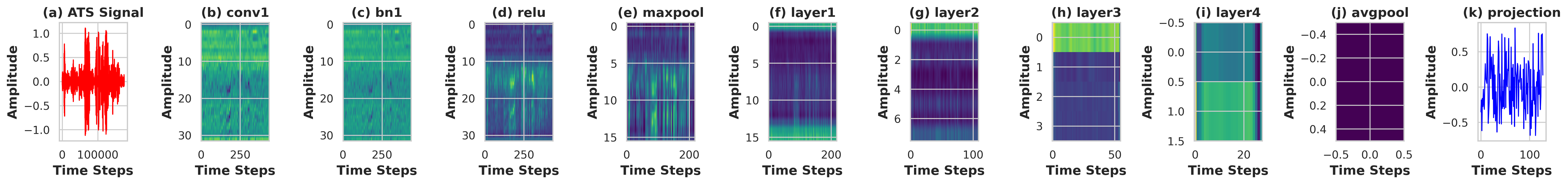} \\
    (b-ii)\includegraphics[width=0.96\textwidth]{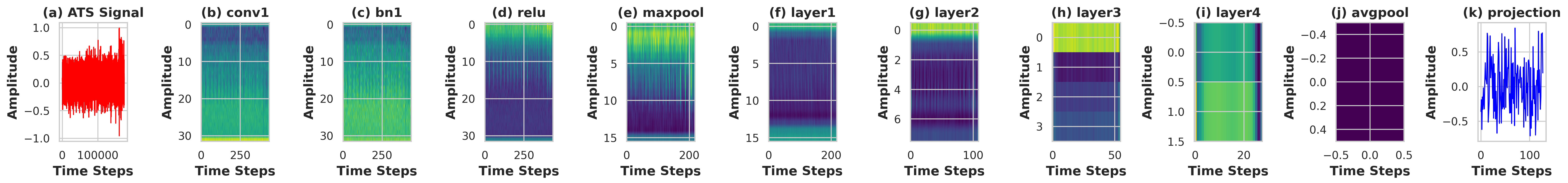} \\
    (b-iii)\includegraphics[width=0.95\textwidth]{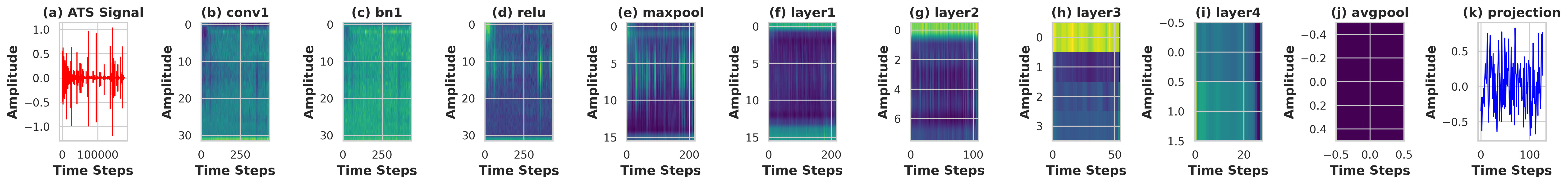}
    \caption{(a) CAM architecture (b) CAM visualization of layers for different ATS signals (i) M (ii) NS (iii) SM.}
    \label{fig:cammodel}
\vspace{-0.6 cm}
\end{figure*} 
\subsection{Statistical Validation}
Statistical validation for inter-class and intra-class distances is done using class centroids and measuring separability between two classes. For each class, the inter-class distance is calculated as
\begin{equation}
    D_{i,j} = || c_i - c_j ||^2,
\end{equation}
where \( D_{i,j} \) is the Euclidean distance between class centroids \( c_i \) and \( c_j \), \( c_i \) and \( c_j \) are the centroids of class \( i \) and class \( j \), respectively. The intra-class compactness measures clustering quality as
\begin{equation}
    d_k (x) = || f_\theta (x) - c_k ||^2,
\end{equation}
where \( d_k(x) \) represents the distance between the feature embedding \( f_\theta(x) \) of sample \( x \) and the centroid \( c_k \) of class \( k \), where \( f_\theta(x) \) is the feature representation of \( x \), and \( c_k \) is the centroid of class \( k \).

\section{Calmness Analysis Model (CAM)}
HRV and CRC are well-established markers of emotional and psychological conditions. Research indicates that auditory stimuli containing rhythmic and harmonic sounds affect the autonomic nervous system and CRC. This research seeks to empirically ascertain if spiritual meditation-based audio has a soothing effect on the human heart compared to music and silence. A custom deep learning model is proposed to categorize recorded ATS signals under three auditory stimuli and examine feature discrepancies using the CAM. 
\vspace{-0.3 cm}
\subsection{Feature Extraction} \label{feat}
To differentiate SM from M and NS, a 25-dimensional feature vector is extracted. Mel frequency cepstral coefficients (MFCCs) briefly overview frequency patterns. In contrast to conventional Fourier features, MFCCs include perceptual weighting, guaranteeing that the most relevant frequency components prevail in the feature space \cite{bhuthegowda2024}. This was very helpful for studying how different ATS can calm people down because it lets us accurately describe changes in the tone, harmonics, and heart resonance caused by specific audio stimuli. MFCC features can be represented as
\begin{equation}
X_{\text{MFCC}} = \left[ MFCC_1, MFCC_2, ..., MFCC_{13} \right] \in \mathbb{R}^{13}.
\end{equation}
The values of those MFCC coefficients can be computed using discrete cosine transform (DCT) from a given audio signal as
\begin{equation}
MFCC_n = \sum_{m=0}^{M} x(m) \cos\left( \frac{\pi}{M} (m + 0.5)n \right).
\end{equation}

Zero-crossing rate (ZCR) and root mean square (RMS) energy are temporal characteristics to assess rhythmic fluctuations and energy distribution \cite{sabry2022healthcare}. Temporal features are computed as
\begin{equation}
X_{\text{Temporal}} = \left[ ZCR, RMS \right] \in \mathbb{R}^{2},
\end{equation}
where $ZCR$ is computed as
\begin{equation}
ZCR = \frac{1}{N-1} \sum_{n=1}^{N-1} \mathbf{1} (w(n)w(n-1) < 0).
\end{equation}
and the $RMS$ energy can be calculated as
\begin{equation}
RMS = \sqrt{\frac{1}{N} \sum_{n=1}^{N} w^2(n)}.
\end{equation}
Fourier transforms assume that signals are stationary, but wavelets allow for multi-resolution analysis, which makes them perfect for biological signals with changing frequency content over time \cite{wang2024}. The mean and standard deviation of the extracted features are calculated to see how the ATS signals are spread out and how they change over time. This gives us a good idea of how SM can calm people down. This selection of features ensures that the CAM fully captures the spectral and temporal aspects of the heart's response to different sounds, leading to a more scientifically sound and clinically useful classification. Wavelet decomposition features are computed as
\begin{equation}
X_{\text{Wavelet}} = \left[ W_1^{\mu}, W_1^{\sigma}, W_2^{\mu}, ..., W_5^{\mu}, W_5^{\sigma} \right] \in \mathbb{R}^{10},
\end{equation}
where each feature can be calculated as
\begin{equation}
W_j^{\mu} = \frac{1}{N} \sum_{i=1}^{N} W_{j,i}, \quad W_j^{\sigma} = \sqrt{\frac{1}{N} \sum_{i=1}^{N} (W_{j,i} - W_j^{\mu})^2}.
\end{equation}

\subsection{Dataset Balancing and Sampling}
If dataset balancing is neglected, imbalance may distort the learning process. It may lead the model to prioritize the dominant class and inadequately reflect minority class tendencies \cite{almadhor2023efficient}. To address this problem, weighted random sampling is applied, assigning a probability weight to each class inversely proportionate to its frequency in the dataset. This ensures the model sees an equal amount of data in all categories. This reduces bias in feature learning and makes it easier for the model to generalize across different heart sound patterns. Mathematically, it can be written as
\begin{equation}
\text{Weight}_i = \frac{\text{Total Samples}}{\text{Samples in Class}_i},
\end{equation}
where Weight${_i}$ denotes the sample weight allocated to each class $i$. 

\subsection{CAM Architecture}
CAM utilizes a bidirectional long short-term memory (BiLSTM) network to analyze ATS signals as it captures sequential dependencies in both forward and backward orientations, which is essential for examining the temporal patterns in CRC variability affected by auditory stimuli shown in Fig.~\ref{fig:cammodel}(a). The hidden state representation $H_T$, has 512 dimensions, integrating outputs from both forward and backward LSTM layers, enhancing the feature extraction process for identifying CRC across audio stimuli. A 25-dimensional feature vector for each ATS signal is extracted as
\[
X = [X_{\text{MFCC}}, X_{\text{Temporal}}, X_{\text{Wavelet}}] \in \mathbb{R}^{25},
\]
Each feature vector \( X \) containing extracted features defined above are processed through a BiLSTM network. The forward pass is represented as
\begin{equation}
\overrightarrow{h_t} = f_{\text{LSTM}}(X_t, \overrightarrow{h_{t-1}}, \overrightarrow{c_{t-1}}).
\end{equation}
Similarly, the backward pass through the network is represented as
\begin{equation}
\overleftarrow{h_t} = f_{\text{LSTM}}(X_t, \overleftarrow{h_{t+1}}, \overleftarrow{c_{t+1}}),
\end{equation}
Finally, the combined impact of forward and backward passes can be represented as
\begin{equation}
H_t = [\overrightarrow{h_t}, \overleftarrow{h_t}] \in \mathbb{R}^{512}.
\end{equation}
The fully connected (FC) layers enhance these representations by using a ReLU activation function to create non-linearity, eliminating vanishing gradients and facilitating complicated feature learning \cite{almadhor2023efficient}. A dropout layer with a 30\% probability is used to reduce overfitting, enhancing the model's generalization to unknown data. The final softmax layer transforms the acquired information into probability distributions across the three target classes, facilitating an interpretable output for classification \cite{lokare2024}. Layer-wise signal transformation is shown in Fig.~\ref{fig:cammodel}(b)- [i, ii, and iii], respectively. Models used in FC layers are represented as
\begin{equation}
H_{\text{fc1}} = \text{ReLU}(W_1 H_t + b_1), \quad H_{\text{fc1}} \in \mathbb{R}^{128}.
\end{equation}
\begin{equation}
H_{\text{drop}} = \text{Dropout}(H_{\text{fc1}}, p=0.3).
\end{equation}
\begin{equation}
y = \text{Softmax}(W_2 H_{\text{drop}} + b_2), \quad y \in \mathbb{R}^{3}.
\end{equation}
Cross-entropy loss is the goal function, and the Adam optimizer is used to improve model convergence \cite{nayef2018effect}. The detailed architecture of the proposed CAM model is shown in Fig.~\ref{fig:cammodel}(a).  The loss function  based on  cross-entropy loss is represented as
\begin{equation}
\mathcal{L} = -\sum_{i=1}^{N} y_i \log(\hat{y}_i).
\end{equation}

\vspace{-0.5 cm}
\subsection{Calmness Evaluation using Mean Analysis \& Pairwise T-tests}
ANOVA is conducted to test if the mean feature values differ significantly across the three groups: $H_0: \mu_{\text{SM}} = \mu_{\text{M}} = \mu_{\text{NS}}$, where $H_a:$ \text{At least one group has a different mean}. The F-statistic is computed as
\begin{equation}
F = \frac{\text{Between-group variance}}{\text{Within-group variance}}.
\end{equation}

If ANOVA reveals a significant difference (\( p < 0.05 \)), pairwise t-tests are conducted to compare each ATS condition as
\begin{equation}
t_{(i,j)} = \frac{\bar{X}_i - \bar{X}_j}{\sqrt{\frac{s_i^2}{N_i} + \frac{s_j^2}{N_j}}},
\end{equation}
where \( \bar{X}_i, \bar{X}_j \) are the means of two groups, \( s_i^2, s_j^2 \) are their variances and \( N_i, N_j \) are their sample sizes.

The calmest category for each feature is determined by identifying the group with the lowest mean value. The overall calmest category is selected using a majority vote as follows:
\vspace{-0.2 cm}
\begin{equation}
C_{\text{calm}} = \arg \max_{C} \sum_{f} \mathbf{1} \left( \min \left( \bar{X}_{f, i}, \bar{X}_{f, j} \right) = C \right),
\end{equation}
\vspace{-0.1 cm}
where \( \bar{X}_{f, i}, \bar{X}_{f, j} \) are the means of feature \( f \) for groups \( i \) and \( j \), respectively. \( \mathbf{1}(\cdot) \) is the indicator function that outputs 1 if the condition inside is true (i.e., the minimum mean of feature \( f \) corresponds to category \( C \)), and 0 otherwise. The sum is taken over all features \( f \), and the overall calmest category \( C_{\text{calm}} \) is the one that maximizes the number of features where the category \( C \) has the lowest mean.

\begin{figure*}[hbt!]
    \centering
    (a) \includegraphics[width=0.85\textwidth]{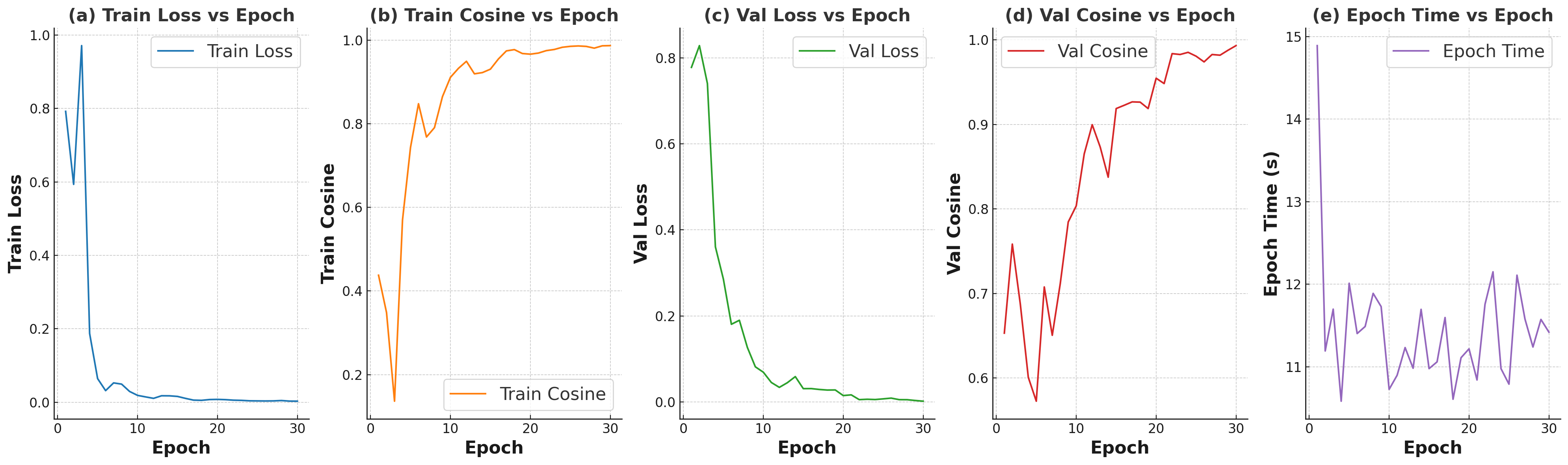} \\
    (b) \includegraphics[width=0.75\textwidth]{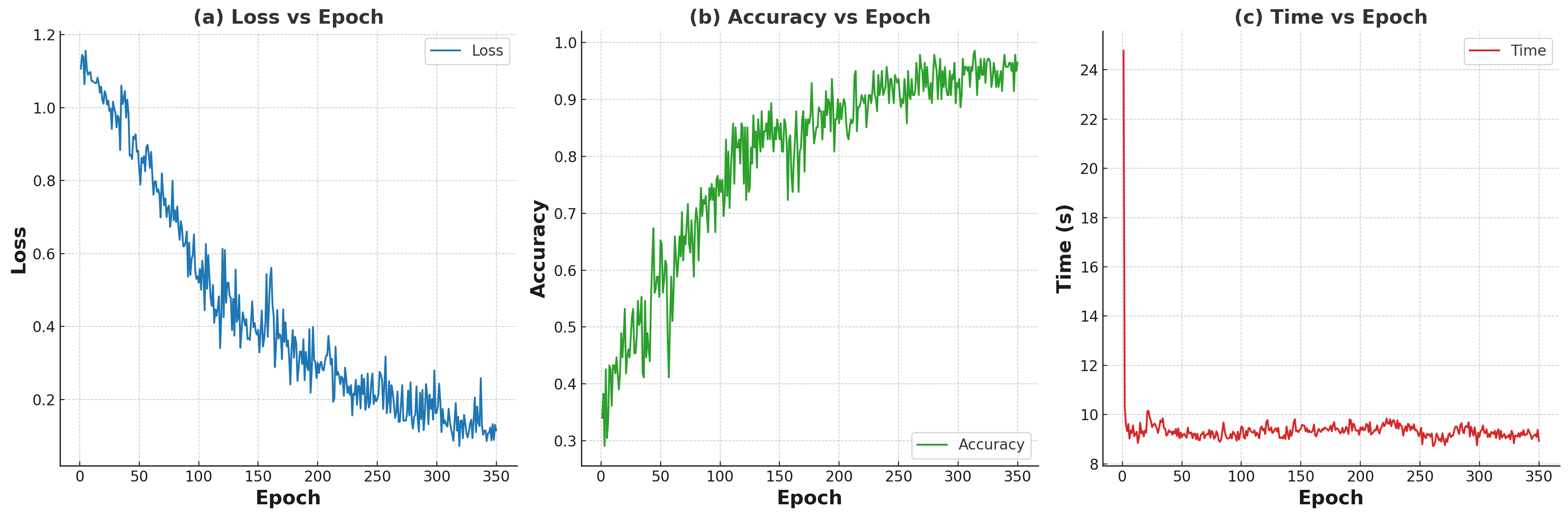}
    \caption{ SMSAT Encoder (a-i) Training loss curves (a-ii) Cosine similarity curves. (b) CAM (i) Loss (ii) Training accuracy (iii) Epoch time}%
    \label{fig:loss}
\vspace{-0.5cm}
\end{figure*} 

\section{Experimental Results and Discussions}
\subsection{SMSAT Performance, Convergence,  Computational Complexity \& Efficiency Analysis}
The self-supervised SMSAT Encoder is trained for 30 epochs, significantly improving performance. The encoders contrastive loss follows an exponential decay
\begin{equation}
\mathcal{L}_{\text{train}}(t) = \alpha e^{-\beta t} + \gamma,
\end{equation}
where $\alpha, \beta$ are constants governing the speed of convergence, and $\gamma$ represents the loss floor. The cosine similarity follows a logistic growth that is
\begin{equation}
C_{\text{val}}(t) = \frac{C_{\max}}{1 + e^{-\lambda (t - t_0)}}.
\end{equation}

At epoch 30, the model achieves: $\mathcal{L}_{\text{train}} = 0.0015, \quad \mathcal{L}_{\text{val}} = 0.0033, C_{\text{train}} = 0.9958, \quad C_{\text{val}} = 0.9932$. These values indicate highly optimized embeddings, minimizing intra-class variance while maximizing inter-class distance.  Fig. \ref{fig:loss}(a)illustrates the training loss and cosine similarity curves.

The total number of parameters in the model is: $P_{\text{total}} = 11,235,904$ with computational complexity: $\mathcal{O}(D \cdot F \cdot M^2)$. $D$ is the input spectrogram dimension, $F$ represents convolutional filters, and $M$ is the fully connected matrix size. The total $FLOPs_{\text{total}} = 200.54 \times 10^6$. The model contains 11.23 million parameters, making it computationally efficient for embedded and real-time applications, and total operations are 200 million, which is manageable on GPU hardware.

\begin{figure*}[!t]
    \centering
    (a) \includegraphics[width=0.97\textwidth]{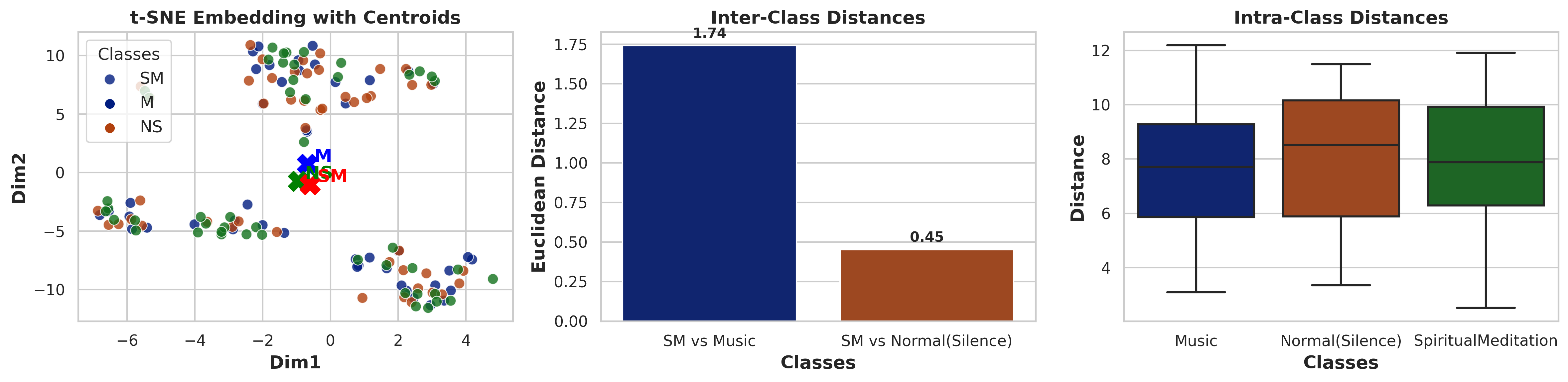} \\
    (b) \includegraphics[width=0.97\textwidth]{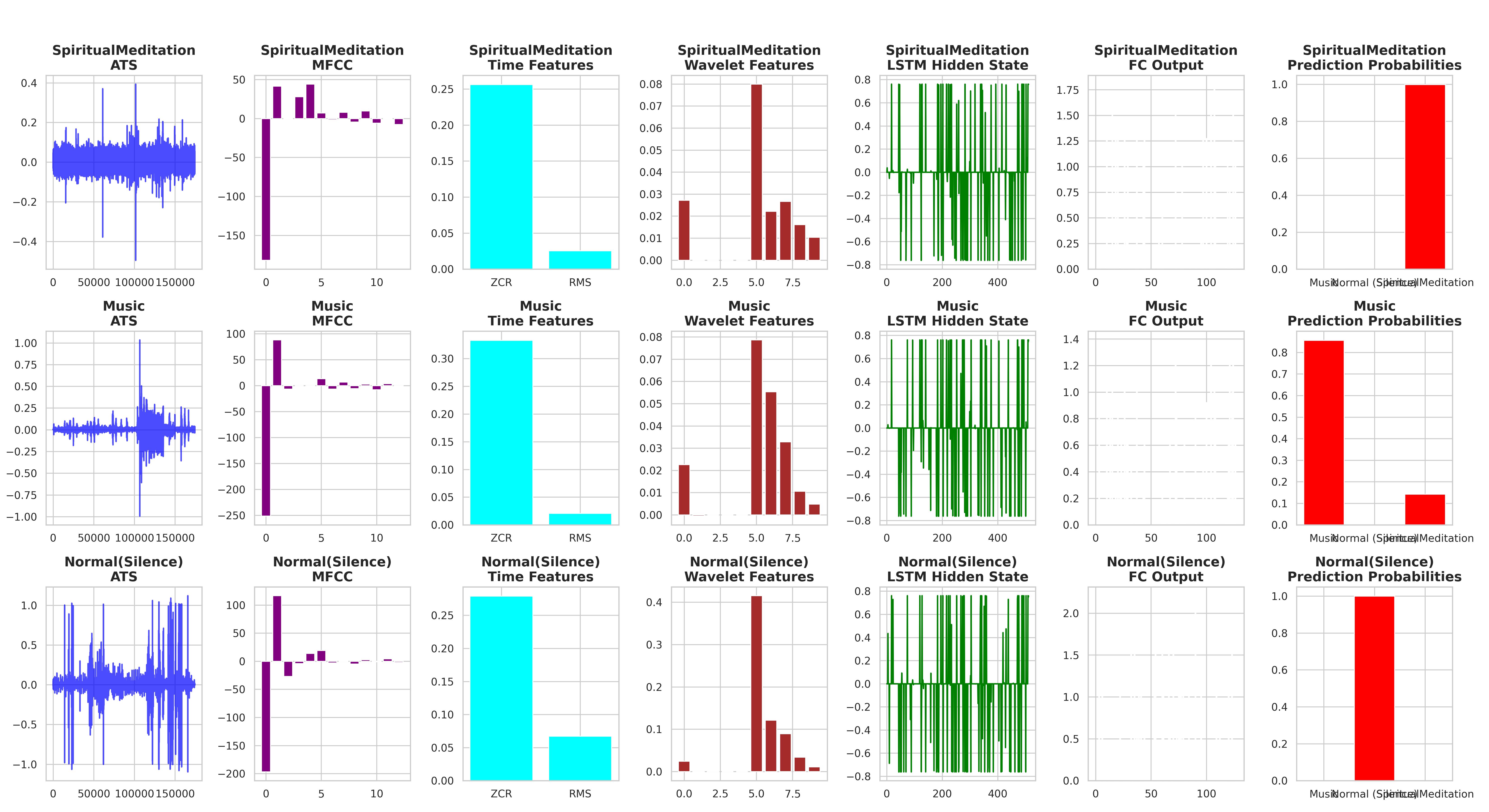}
        \caption{ (a) SMSAT (i) t-SNE Embedding with class centroids (ii) Inter-class distances  (iii) Intra-class distances (b) CAM visualization of all ATS signals.}
    \label{fig:model_visualization}
\vspace{-0.7 cm}
\end{figure*}

Now, we evaluate the SMSAT feature space analysis and class separability. 
For inter-class distance: $d_{\text{SM, M}} \&= 1.846$ and  $d_{\text{SM, NS}} \&= 0.214$. This indicates that SM and NS share more common features, making classifying these two categories more challenging. Intra-class compactness suggests that all three classes exhibit similar internal variance, with SM having the most compact distribution $
\bar{d}_{\text{SM}} \&= 9.58, \quad \bar{d}_{\text{Music}} = 9.51, \quad \bar{d}_{\text{NS}} = 9.65
$. Class separability evaluation confirms that SM is distinct from M but closely related to NS $S_{\text{SM, M}} \&= 0.62, S_{\text{SM, NS}} \&= 0.06 $. The t-SNE embedding is calculated using as  
\begin{equation}
C = \sum_{i \neq j} p_{ij} \log \frac{p_{ij}}{q_{ij}},
\end{equation}
where $p_{ij}, q_{ij}$ represents probability distributions in high and low-dimensional spaces. The results demonstrate the effectiveness of self-supervised learning and the proposed encoder. High cosine similarity  ($\sim 0.99$) confirms that the SMSAT encoder has learned stable audio embeddings, as shown in Fig. \ref{fig:model_visualization}(a). Significant inter-class distance between SM and M suggests effective feature extraction. The low separation between SM and NS highlights acoustic similarities, which could impact classification performance Fig. \ref{fig:model_visualization}(b). Computational efficiency ensures feasibility for real-time deployment, with 11.2M parameters and 200M FLOPs.

The CAM model hyper-parameters: batch size is 512, 350 epochs, 0.005 learning rate, and 30\% dropout rate. Model evaluation and performance metrics are calculated using accuracy and a confusion matrix. Fig.~\ref{fig:model_visualization}(b) presents a comprehensive visualization of how the CAM processes different audio classes. Each row in the figure represents an audio sample from a distinct category, and each column displays the transformation of the input signal to varying stages of the model's feature extraction and classification process.

The training of the CAM shows a variable but overall promising trend in accuracy, loss, and epoch time. Initially, the model exhibits a relatively low accuracy of 34.75\% and a loss value of 1.0925 at epoch 1. However, as training progresses, we observe an upward trajectory in accuracy (↑), reaching 35.46\% by the 5th epoch, suggesting a slight but consistent improvement in performance. Meanwhile, the loss decreases (↓) from 1.0925 to 1.0923, indicating a small yet steady reduction in error. Epoch time varies significantly, with a noticeable fluctuation between 9 and 25 seconds, with no clear, consistent trend, which could indicate the complexity or variability of the model's learning process at different stages. This epoch time variability could be optimized further as the model stabilizes its accuracy and loss rates over extended training periods, as shown in Fig.~\ref{fig:loss}(b).  

CAM is trained with a learning rate of 0.005, a relatively modest yet practical value for controlling the speed of the model's convergence. 350 epochs allow sufficient time for the model to learn and improve its performance over multiple iterations. The batch size of 512 ensured that the model could efficiently process a large amount of data at each step, balancing speed and stability during training. The input dimension is set to 25, which likely corresponds to the number of features in the input data. In contrast, the hidden dimension of 256 indicates a moderate level of model complexity for capturing patterns in the data. A dropout rate of 0.3 was implemented to prevent overfitting, encouraging the model to generalize better. The model utilized the Adam optimizer, known for its efficiency in adapting learning rates during training, and the cross-entropy loss criterion, which is suitable for classification tasks. The total GPU training time amounted to 3436.82 seconds, demonstrating the model's efficient utilization of P100 GPU during the training process. 

The comparison of the train and test classification results reveals key insights into the model's performance as shown in Table\ref{table:calmness-ans}. For Music, the training precision, recall, and F1-score were 96\%, 98\%, and 97\%, respectively, suggesting that the CAM is proficient in identifying music with low false positives. However, the test results show perfect performance (99.99\%) across all metrics, indicating a substantial improvement in test accuracy (↑). For NS, the training performance achieved precision and recall of 98\% and 99.99\%, respectively, and an F1 of 99\%. SM showed perfect precision (99.99\%) during both training and testing, but the recall dropped from 96\% during training to 99.99\% during testing, leading to a substantial rise in the F1 from 98\% to 99.99\% (↑). This indicates that the model could capture this class during the test phase better, eliminating any earlier recall issues. As shown in Table \ref{tab:trr}, the training and testing results for accuracy, precision, recall, and F1-score demonstrate high performance across all classes (M, NS, SM) and the highest among all is SM. 
\begin{table}[!b]
\centering
\scriptsize
\caption{CAM Train and Test Results}
\label{tab:trr}
\begin{tabular}{ |l|l|l|l|l|l| }
  \hline
  \textbf{Metrics} & \textbf{Train M} & \textbf{Train NS} & \textbf{Train SM} & \textbf{Test M} & \textbf{Test NS} \\
  \hline
  \textbf{Accuracy} & 98\% & 100\% & 96\% & 100\% & 100\% \\
  \hline
  \textbf{Precision} & 96\% & 98\% & 99.99\% & 99.99\% & 99.99\% \\
  \hline
  \textbf{Recall} & 98\% & 99.99\% & 96\% & 99.99\% & 99.99\% \\
  \hline
  \textbf{F1-Score} & 97\% & 99.99\% & 98\% & 99.99\% & 99.99\% \\
  \hline
\end{tabular}
\end{table}

\begin{table*}[!t]
\centering
\scriptsize
\caption{Feature-wise Calmness Analysis of All the Classes.}
\label{table:calmness-ans}
\renewcommand{\arraystretch}{1.1}
\begin{tabular}{l|c|c|c|c|c|c|c|c|c|c}
\toprule
\textbf{Feature} & \multicolumn{3}{c|}{\textbf{Mean}} & \textbf{Mean Comparison} & \textbf{Calmest Class} & \textbf{ANOVA p-value} & \multicolumn{3}{c|}{\textbf{Pairwise p-values}} & \textbf{Result} \\
\cline{2-4} \cline{8-10}
 & \textbf{SM} & \textbf{NS} & \textbf{M} &  & &  & \textbf{SM vs M} & \textbf{SM vs NS} & \textbf{M vs NS} & \\
\midrule
MFCC\_0  & \textbf{-254.2791} & -249.5712 & -245.2362 & SM $<$ NS $<$ M & SM & 0.8719 & 0.5749 & 0.7915 & 0.8051 & No diff \\
MFCC\_1  & 93.7285 & \textbf{87.9650} & 99.2803 & SM $>$ NS $<$ M & NS & 0.6528 & 0.6550 & 0.6439 & 0.3522 & No diff \\
MFCC\_2  & \textbf{-6.4426} & -9.1582 & -3.7299 & SM $>$ NS $<$ M & NS & 0.7109 & 0.6854 & 0.6800 & 0.4101 & No diff \\
MFCC\_3  & 12.7264 & \textbf{9.7897} & 12.8667 & SM $>$ NS $<$ M & NS & 0.3894 & 0.9579 & 0.2322 & 0.2250 & No diff \\
MFCC\_4  & \textbf{11.3003} & 11.4780 & 12.4645 & SM $<$ NS $<$ M & SM & 0.8371 & 0.5949 & 0.9315 & 0.6333 & No diff \\
MFCC\_5  & \textbf{12.1020} & 12.6197 & 13.8407 & SM $<$ NS $<$ M & SM & 0.4863 & 0.2338 & 0.7289 & 0.4145 & No diff \\
MFCC\_6  & -3.6092 & \textbf{-4.4564} & -2.9514 & SM $>$ NS $<$ M & NS & 0.5839 & 0.6441 & 0.5689 & 0.3095 & No diff \\
MFCC\_7  & 4.6322 & \textbf{4.2813} & 4.4587 & SM $>$ NS $<$ M & NS & 0.9222 & 0.8536 & 0.6811 & 0.8315 & No diff \\
MFCC\_8  & \textbf{-3.0367} & -2.5192 & -3.3189 & SM $<$ NS $>$ M & M & 0.6412 & 0.7593 & 0.5451 & 0.3364 & SM vs M diff \\
MFCC\_9  & \textbf{3.4204} & 4.5822 & 3.6432 & SM $<$ NS $>$ M & SM & 0.4880 & 0.8399 & 0.2518 & 0.3460 & No diff \\
MFCC\_10 & \textbf{-3.4915} & -3.0753 & -3.0256 & SM $<$ NS $<$ M & SM & 0.7996 & 0.5728 & 0.5751 & 0.9453 & No diff \\
MFCC\_11 & \textbf{1.7940} & 2.9238 & 2.0206 & SM $<$ NS $>$ M & SM & 0.3404 & 0.7905 & 0.1718 & 0.2473 & No diff \\
MFCC\_12 & \textbf{-3.5625} & -2.8183 & -3.6268 & SM $<$ NS $>$ M & M & 0.4586 & 0.9277 & 0.3338 & 0.2452 & SM vs M diff \\
ZCR      & 0.2739 & 0.2693 & \textbf{0.2480} & SM $>$ NS $>$ M & M & 0.5760 & 0.3272 & 0.8625 & 0.4141 & M vs SM diff \\
RMS      & 0.0596 & \textbf{0.0591} & 0.0644 & SM $>$ NS $<$ M & NS & 0.9163 & 0.7445 & 0.9747 & 0.6906 & No diff \\
Wavelet\_0 & \textbf{-254.2791} & -249.5712 & -245.2362 & SM $<$ NS $<$ M & SM & 0.8719 & 0.5749 & 0.7915 & 0.8051 & No diff \\
Wavelet\_1 & 93.7285 & \textbf{87.9650} & 99.2803 & SM $>$ NS $<$ M & NS & 0.6528 & 0.6550 & 0.6439 & 0.3522 & No diff \\
Wavelet\_2 & \textbf{-6.4426} & -9.1582 & -3.7299 & SM $>$ NS $<$ M & NS & 0.7109 & 0.6854 & 0.6800 & 0.4101 & No diff \\
Wavelet\_3 & 12.7264 & \textbf{9.7897} & 12.8667 & SM $>$ NS $<$ M & NS & 0.3894 & 0.9579 & 0.2322 & 0.2250 & No diff \\
Wavelet\_4 & \textbf{11.3003} & 11.4780 & 12.4645 & SM $<$ NS $<$ M & SM & 0.8371 & 0.5949 & 0.9315 & 0.6333 & No diff \\
Wavelet\_5 & \textbf{12.1020} & 12.6197 & 13.8407 & SM $<$ NS $<$ M & SM & 0.4863 & 0.2338 & 0.7289 & 0.4145 & No diff \\
Wavelet\_6 & -3.6092 & \textbf{-4.4564} & -2.9514 & SM $>$ NS $<$ M & NS & 0.5839 & 0.6441 & 0.5689 & 0.3095 & No diff \\
Wavelet\_7 & 4.6322 & \textbf{4.2813} & 4.4587 & SM $>$ NS $<$ M & NS & 0.9222 & 0.8536 & 0.6811 & 0.8315 & No diff \\
Wavelet\_8 & \textbf{-3.0367} & -2.5192 & -3.3189 & SM $<$ NS $>$ M & M & 0.6412 & 0.7593 & 0.5451 & 0.3364 & SM vs M diff \\
Wavelet\_9 & \textbf{3.4204} & 4.5822 & 3.6432 & SM $<$ NS $>$ M & SM & 0.4880 & 0.8399 & 0.2518 & 0.3460 & No diff \\
\bottomrule
\end{tabular}
\end{table*}

\vspace{-0.2cm}
\subsection{Calmness Evaluation using Pairwise t-tests}
The details of each feature are discussed in section \ref{feat}. A statistical hypothesis testing approach is used to evaluate the effect of different auditory stimuli on CRC. We compare SM, M, and NS using statistical measures to quantify the calmness effect. The mean characteristic values, ANOVA p-values and the determination of the calmest category are presented in Table \ref{table:calmness-ans}. 
\vspace{-0.1 cm}
\subsubsection{SM vs. NS} Silence represents as a baseline state, and ideally, a calming auditory stimulus should minimize deviations from this natural state. Mathematically, based on the mean characteristic values and the results of Table \ref{table:calmness-ans}, \textbf{$X_Q \approx X_N < X_M$} suggest that SM and NS are closer in terms of CRC than M, indicating a potential calming effect. The low ANOVA p values for features such as RMS/ZCR indicate that significant differences were observed between SM and M, which reinforces that SM and silence are more aligned in their impact on CRC.

\subsubsection{SM vs. M} Statistical comparisons reveal that M introduces more fluctuations in CRC than SM. Comparison of pairwise ZCR and RMS Energy, shows \textbf{$X_Q < X_M$}. As inferred from the Table \ref{table:calmness-ans}, the ANOVA results support the hypothesis that SM is a more calm auditory stimuli than Music. 

Statistical evidence supports the conclusion that SM is significantly closer to NS than M in terms of their effects on CRC. From the pairwise t tests and the ANOVA results, we can confidently conclude that \textbf{$\text{Silence (Normal)} \approx \text{Spiritual Meditation} < \text{Music}$}. Thus, SM provides a more stable and calming experience than M.

\begin{itemize}
    \item \textbf{ANOVA Results}: The ANOVA p-values for most features, including RMS and ZCR, are above 0.05 (e.g., p = 0.9163 for RMS, p = 0.5760 for ZCR), indicating that SM and NS exhibit similar effects, while M introduces greater variation.
    \item \textbf{Pairwise t-test Comparisons}: The pairwise comparisons revealed that the most significant difference between SM and M is in features such as RMS and ZCR. SM consistently shows smaller fluctuations than M, confirming its calming effect closer to Silence.
\end{itemize}

In conclusion, \textbf{SM} exhibits a calming effect on CRC, aligning more closely with NS than M, which introduces more significant fluctuations and excitation.
\vspace{-0.3 cm}
\subsection{Comparison with the State-of-the-art Schemes}
Table \ref{tab:AI_models_comparison} provides a fair comparison of AI models from various studies with the proposed model. From the table, it can be clearly noticed that the proposed model consistently outperforms other models in terms of accuracy, achieving a remarkable 99\% accuracy with both the SMSAT ATS Encoder and the CAM Model. In contrast, the highest accuracy reported in the literature is 90\% by the model from \cite{zawad2024stress}. The proposed model's accuracy surpasses the current state-of-the-art methods by a significant margin, highlighting its potential for superior SM detection. This improvement in performance underscores the efficacy and robustness of the proposed approach when compared to traditional models.
 
\begin{table}[h]
\centering
\caption{Comparison of the Proposed Model with the State-of-the-art Schemes.}
\resizebox{\columnwidth}{!}{
\begin{tabular}{|l|l|l|l|}
\hline
\textbf{Paper \& Year} & \textbf{Dataset} & \textbf{AI Model} & \textbf{Accuracy} \\ \hline
\cite{palma2023stress} 2023 & Stress Detection Levels & CNN, LSTM & 85\% \\ \hline
\cite{almadhor2023efficient} 2023 & Stress Detection based on Electrodermal Activity & Stacking model & 86\% \\ \hline
\cite{masri2023mental} 2023 & Mental stress levels at the workplace & Random Forest, SVM & 84\% \\ \hline
\cite{hemakom2024ecg} 2024 & ECG and EEG data from various individuals & CNN, SVM & 88\% \\ \hline
\cite{zawad2024stress} 2024 & Stress detection using HRV & CNN, RNN & 90\% \\ \hline
\cite{di2024anxiety} 2024 & Anxiety detection from physiological signals & CNN, ANN & 87\% \\ \hline
Proposed 2025 & SMSAT (NS Vs M Vs SM) & SMSAT ATS Encoder & 99\% \\ \hline
Proposed 2025 & SMSAT (NS Vs M Vs SM) & CAM Model & 99\% \\ \hline
\end{tabular}
}
\vspace{-0.5 cm}
\label{tab:AI_models_comparison}
\end{table}

\section{Conclusion}
In this paper, we presented a comprehensive multimodal investigation into the affective and physiological effects of spiritual meditation (SM) in comparison to music (M) and natural silence (NS). Leveraging a newly developed and demographically diverse dataset, SMSAT, we provided empirical evidence demonstrating that SM induces physiological responses that are statistically indistinguishable from natural resting states, while significantly differing from those elicited by music. To support this analysis, we proposed a contrastive learning-based SMSAT audio encoder and a deep learning classification framework, the Calmness Analysis Model (CAM), both of which achieved near-perfect accuracy in discriminating among auditory conditions based on acoustic time-series features. These models enabled fine-grained assessment of affective states and physiological responses, setting a new technical benchmark for affective computing in auditory-based interventions. Our findings suggest that SM represents a highly effective, non-invasive modality for stress reduction and emotional regulation, with potential applications in mental health diagnostics, personalized therapy, and emotion-aware technologies. Future work will explore the longitudinal impact of SM, generalization to diverse clinical populations, and integration into real-time affective computing systems.
\vspace{-0.2 cm}

\bibliographystyle{IEEEtran}
\bibliography{cas}

\end{document}